\documentclass[twocolumn,showpacs,preprintnumbers,amsmath,amssymb]{revtex4}

\usepackage{mathrsfs}
\usepackage{amsmath}
\usepackage{graphicx}
\usepackage{subfigure}
\usepackage{bm}
\usepackage{epstopdf}
\usepackage{float}
\usepackage{hyperref}
\usepackage{color}
\usepackage{dcolumn}
\usepackage{bm}
\usepackage{graphicx}

\usepackage{hyperref}

\newcommand{\bea}{\begin{eqnarray}}
\newcommand{\eea}{\end{eqnarray}}
\newcommand{\beq}{\begin{equation}}
\newcommand{\eeq}{\end{equation}}
\newcommand{\nn}{\nonumber}

\def\/{\over}

\begin{document}
\title{\bf Resonance interaction energy between two accelerated identical atoms in a coaccelerated frame and the Unruh effect}
\author{Wenting Zhou$^{1,2}$,  Roberto Passante$^{2,3}$, and Lucia Rizzuto$^{2,3}$}
\address{$^{1}$ Center for Nonlinear Science and Department of Physics, Ningbo
University, Ningbo, Zhejiang 315211, China}
\address{$^2$ Dipartimento di Fisica e Chimica, Universit\`{a} degli Studi di Palermo and CNISM, Via Archirafi 36, I-90123 Palermo, Italy}
\email{roberto.passante@unipa.it}
\address{$^3$ INFN, Laboratori Nazionali del Sud, I-95123 Catania, Italy}

\begin{abstract}
We investigate the resonance interaction energy between two uniformly accelerated identical atoms, interacting with the scalar field or the electromagnetic field in the vacuum state, in the reference frame coaccelerating with the atoms. We assume that one atom is excited and the other in the ground state, and that they are prepared in their correlated symmetric or antisymmetric state. Using perturbation theory, we separate, at the second order in the atom-field coupling, the contributions of vacuum fluctuations and radiation reaction field to the energy shift of the interacting system. We show that only the radiation reaction term contributes to the resonance interaction between the two atoms, while Unruh thermal fluctuations, related to the vacuum fluctuations contribution, do not affect the resonance interatomic interaction. We also show that the resonance interaction  between two uniformly accelerated atoms, recently investigated in the comoving (locally inertial) frame, can be recovered in the coaccelerated frame, without the additional assumption of the Fulling-Davies-Unruh temperature for the quantum fields (as necessary for the Lamb shift, for example). This indicates, in the case considered, the equivalence between the coaccelerated frame and the locally inertial frame.
\end{abstract}
\pacs{04.62.+v, 03.70.+k, 42.50.Lc}
\maketitle

\section{Introduction}

Quantum field theory in accelerated frames predicts that a detector (an atom, for example) uniformly accelerated in the Minkowski vacuum perceives the vacuum state as a thermal bath at a temperature proportional to its acceleration, $T={a/2\pi}$, $a$ being the atomic acceleration \cite{Fulling73,Davies75,Unruh76}.  In qualitative terms this phenomenon, known as the Unruh or Unruh-Fulling-Davies effect, originates from the time-dependent Doppler shift of the quantum vacuum field detected by the accelerated atom/detector \cite{AM04}. A fundamental consequence of this effect is that the concept of  {\em particle} in quantum field theory is essentially observer-dependent.  Initially treated as a purely kinematic effect, the Unruh effect has stimulated intense theoretical investigations on the dynamical properties of uniformly accelerated systems, for example the emission of radiation from uniformly accelerated atoms \cite{CHM08}. The importance of the Unruh effect is nowadays recognized, not only in its own right, but also in connection with many other actual research topics, such as black hole evaporation \cite{Hawking74}, cosmological horizon \cite{GH77}, and also  applications in quantum information science \cite{KY03,LM09}.
Several experimental schemes have been proposed to detect this very tiny effect, intertwining quantum physics and general relativity, in the laboratory, but it has not been yet observed. In fact, it is necessary to reach accelerations of the order of $\sim 10^{20}m/s^2$ to obtain an Unruh radiation corresponding to a temperature of a few Kelvin \cite{BL83,CT99,BMBCE08,MFM11}. In this context, it has been recently argued that van der Waals/Casimir-Polder interatomic interactions between two accelerating atoms could be good candidates for an indirect detection of the Unruh effect at experimentally reasonable accelerations \cite{NP13,MNP14}.

Phenomena closely related to the Unruh effect are the dynamical Casimir effect, that is the emission of electromagnetic radiation from a single accelerated mirror in the vacuum \cite{Moore70,Dodonov09} and the dynamical Casimir-Polder interactions, which originate from a nonadiabatic change of some physical parameter of the system (such as the atomic transition frequency, or the dielectric properties of a physical boundary) \cite{VP08,MPV10,ABCNPRRS14}.
Both the Unruh effect and the dynamical Casimir and Casimir-Polder effects stress the nontrivial nature of the quantum vacuum, underlining  its intrinsic dynamical structure.

Although the Unruh effect has been extensively investigated up to now from a theoretical point of view, controversies concerning with the real existence of this effect and its physical meaning remain open \cite{FNMG09, BS13}.  On the other hand, it has been recently argued that the existence of the Unruh effect is mandatory for the consistency of quantum field theory \cite{VM01}. It is therefore relevant to investigate theoretically all physical manifestations of the Unruh effect in different physical systems, as well as possible experimental setups to detect this elusive phenomenon at the boundary between quantum mechanics and general relativity.
Recently, radiative properties of atoms in noninertial motion \cite{NP13,AM94,AM95a,AM95b,Passante98,Rizzuto07, RS09a,RS09b,RS11,MNPRS14, ZYL06,YZ06,ZY07, ZY10} or atoms at rest immersed in a thermal bath \cite{TC03,ZY09,SYZ10,ZYW12}, have been investigated, also aiming for proposals of experimental verifications of the Unruh effect. The main aim of these investigations is also to explore the effect of a uniform acceleration on the dynamical properties of atomic systems, and at which extent the Unruh equivalence of acceleration and temperature is valid.
In this context, many investigations in the literature have been concerned with the question of the equivalence between the locally inertial- and coaccelerated-frame points of view.
In particular, it has been shown in different specific cases that a complete agreement between the physical results obtained in the comoving (locally inertial) and coaccelerated frame can be restored, providing to assume an Unruh temperature proportional to the acceleration of the system, for the quantum fields in the Rindler frame; examples are the excitation rate of a uniformly accelerated atom \cite{ZY07,ZY08}, the rate of photon emission with a given transverse momentum \cite{HMS92a,HMS92b} and the decay of accelerated protons \cite{VM01}.  From the previous considerations, it seems that the existence of the Unruh effect is a requirement for the compatibility of physical results in different (locally inertial and coaccelerated) frames.  A question naturally emerges, that is if this assumption is a necessary condition for any radiative process, or if some effect exists that is observer-independent, for which Minkowski and Rindler observers give the same predictions, without any additional assumption of an Unruh temperature.

In this paper, we explore this issue investigating the resonance interaction energy between two uniformly accelerated atoms, from the point of view of a coaccelerated observer. The resonance interaction between uniformly accelerated atoms in the Minkowski vacuum has been recently investigated in the comoving frame \cite{RLMNSZP16}. Also, the radiative properties of uniformly accelerated entangled atoms have been recently investigated \cite{MS16}. Here, we consider the resonance energy between two atoms prepared in a {\em Bell-type} state in the Rindler spacetime, with also the aim to discuss the equivalence of the results obtained in the Rindler frame with those obtained in a locally inertial frame.

Resonance interactions between atoms occur when one or both atoms are in an excited state and an exchange of real or virtual photons between the two atoms is involved. These interactions are usually fourth-order effects in perturbation theory.
However, if the two identical atoms are  prepared in a correlated (symmetric or antisymmetric) state, with one atom in its ground state and the other in an excited state, the excitation is delocalized among the two atoms and the resonance interaction is a second-order effect in the atom-field coupling constant \cite{CT98,Salam10}. This stems from the fact that, in the symmetric or antisymmetric state, the two atomic dipoles are correlated, while for factorized states (as in the case of dispersion interaction, for example) they must be correlated by vacuum fluctuations \cite{PPR03,RPP07}. For this reason this interaction can be much stronger than the resonance interaction between atoms in factorized states, even if it requires preparing the system in a correlated state. Resonance interactions are very long range effects, scaling as $r^{-3}$ ($r$ being the interatomic distance) in the limit of short distance (compared with the atomic transition wavelength $(r\ll\lambda_0)$), and as $r^{-1}$ for $r\gg\lambda_0$. This behavior should be compared with the usual dispersion interactions between ground-state atoms, that scales as $r^{-7}$ in the far-zone limit, or the interaction between atoms in excited factorized states that in the same limit scales as $r^{-2}$.

Recently, it has been discussed that resonant interactions can be modified (enhanced or inhibited) in various circumstances, for example when the atoms are immersed in a structured environment such as a photonic crystal or a waveguide \cite{IFTPRP14,NPR16}. The relation of the resonance interaction with other physical processes, such as the resonant energy transfer \cite{Salam10}, as well as its role in some coherent biological processes \cite{PP13}, have been also investigated.
Also, very recently the resonance interaction between two uniformly accelerated atoms in the Minkowski space has been investigated from the point of view of a comoving frame \cite{RLMNSZP16}. It has been shown that Unruh thermal fluctuations do not affect the resonance interaction between the two accelerated atoms, which is exclusively due to the radiation reaction field. Nevertheless, the noninertial motion of the atoms affects the resonant interatomic energy, showing new properties, ultimately related to the peculiar structure of the quantum vacuum of the electromagnetic field in a locally inertial frame \cite{RLMNSZP16}. In particular, nonthermal effects of the atomic acceleration, related to the noninertial character of accelerated motion, result in a different scaling with the distance and a different dependence on the acceleration compared to those expected from the known Unruh acceleration-temperature equivalence.
Here we focus our attention on the resonance interaction between atoms in non-inertial motion, from the point of view of a coaccelerated observer, and discuss the relation between the results obtained in the comoving and coaccelerated frames. Following the procedure of Dalibard \emph{et al.} \cite{DDC82, DDC84} to separate vacuum fluctuations and radiation reaction contributions, we calculate the resonance interaction between two atoms prepared in a correlated state, and interacting with the scalar or the electromagnetic field in the vacuum state (in the Rindler frame). In agreement with the results in \cite{RLMNSZP16}, we show that vacuum field fluctuations do not affect the resonance interaction between the two correlated atoms, which is exclusively due to the radiation reaction contribution.
In both cases considered (scalar and electromagnetic field) we obtain in the coaccelerated frame the same expressions of the resonance interaction obtained in the comoving frame, without any additional assumption of an Unruh temperature for the quantum field. This behavior is basically related to the {\em nonthermal} feature of the resonance interaction.
Therefore we find that, at variance of other physical effects such as the Lamb shift or the rate of spontaneous emission of uniformly accelerated atoms, the resonance interaction between two accelerated atoms is observer-independent.
The equivalence obtained shows that the Unruh thermal bath, although present in the comoving frame of the accelerated atoms, does not play any role in some specific radiative processes such as the resonance interaction between accelerated atoms.

The paper is organized as follows. In Sec. \ref{sec:2}  we introduce our model and calculate the resonance interaction between two  correlated
accelerated identical atoms, interacting with the scalar field in the vacuum state, in the coaccelerated frame; we then compare our results with those obtained in the locally inertial frame. In Sec. \ref{sec:3} we extend our results to the case of accelerated atoms
interacting with the electromagnetic field in the vacuum state, using the Weyl gauge, comparing the results obtained in the coaccelerated frame with previous results obtained in the comoving frame and discuss their physical meaning. Section \ref{sec:4} is finally devoted to our conclusive remarks.

\section{The scalar field case}
\label{sec:2}

We consider two identical two-level atoms, labeled with $A$ and $B$,  uniformly accelerated along parallel trajectories, in the Minkowski vacuum. We suppose that the two atoms accelerate with the same acceleration $a$ along the $x$ direction, while their distance is along $z$; denoting the coordinates in an inertial frame with $(t,x,y,z)$, the trajectories of the two atoms are given by
\bea
&\ &t(\tau)={1\/a}\sinh{a\tau},\,\,\, x_{A/B}(\tau)={1\/a}\cosh{a\tau},\nonumber\\
&\ &y_{A/B}(\tau)=0, \,\,\,\ z_A(\tau)=z_1,\,\,\, z_B(\tau)=z_2,
\label{eq:1}
\eea
where $a$ is the proper acceleration (we use units such that $k_B=\hbar=c=1$; in these units the acceleration has the same dimension of the temperature).

The inertial frame can be related to the coaccelerated frame by the following coordinate transformation,
\beq
t(\tau,\xi)={1\/a}e^{a\xi}\sinh{a\tau}\;,\quad\;x(\tau,\xi)={1\/a}e^{a\xi}\cosh{a\tau}\;
\label{eq:2}
\eeq
with $(\tau,\xi,y,z)$ describing the coordinates in the coaccelerated frame. Then the Minkowski spacetime with the metric
\beq
ds^2=dt^2-dx^2-dy^2-dz^2\;,
\label{eq:3}
\eeq
is correspondingly transformed to the right Rindler wedge
\beq
R_{+}=\{|t|\leq x\}\;,
\label{eq:4}
\eeq
the metric of which can be described by
\beq
ds^2=e^{2a\xi}(d\tau^2-d\xi^2)-dy^2-dz^2\;.
\label{eq:5}
\eeq
An atom with positive acceleration $a$ makes only the right Rindler wedge $R_{+}$ completely accessible to a coaccelerated observer~\cite{Boulware80}.

The scalar field operator in the Rindler wedge can be expanded in terms of a complete set of scalar field modes as
\begin{eqnarray}
& & \phi(\tau,\vec{x})=\int^{\infty}_0d\omega\int^{\infty}_{-\infty}dk_y\int^{\infty}_{-\infty}dk_z \nonumber\\
& &\times\biggl [b_{\omega,k_y,k_z}(\tau)v_{\omega,k_y,k_z}(\vec{x})+b^{\dag}_{\omega,k_y,k_z}(\tau)v^{\star}_{\omega,k_y,k_z}(\vec{x})\biggr],
\label{eq:6}
\end{eqnarray}
with $b^{\dag}_{\omega,k_y,k_z}(\tau)$ ($b_{\omega,k_y,k_z}(\tau)$) the creation (annihilation) bosonic operators of Rindler particles at the proper time $\tau$, and
\begin{eqnarray}
&\ &v_{\omega,k_y,k_z}(\vec{x})={1\/2\pi^2\sqrt{a}}\sqrt{\sinh\biggl({\pi\omega\/a}\biggr)}K_{i{\omega\/a}} \biggl({1\/a}k_{\perp}e^{a\xi}\biggr)\nonumber\\
&\ &\times e^{ik_yy+ik_zz},
\label{eq:7}
\end{eqnarray}
with $k_{\perp}=\sqrt{k_y^2+k_z^2}$. The Hamiltonian of the scalar field is then expressed in terms of the annihilation and creation operators as
\beq
H_F(\tau)=\int^{\infty}_0 d\omega\int^{\infty}_{-\infty} d k_y\int^{\infty}_{-\infty} d k_z\;\omega\;b^{\dag}_{\omega,k_y,k_z}(\tau)b_{\omega,k_y,k_z}(\tau)\;.
\label{eq:8}
\eeq
We now suppose that the two identical atoms are linearly and locally coupled to the relativistic massless scalar field. The Hamiltonian of the interacting system assumes then the form
\begin{eqnarray}
&\ &H(\tau)=\omega_0\left(\sigma^A_3(\tau)+\sigma^B_3(\tau)\right)+H_F(\tau)\nonumber\\
&\ &+\lambda(\sigma^A_2(\tau)\phi(x_A(\tau))+\sigma^B_2(\tau)\phi(x_B(\tau))) ,
\label{eq:9}
\end{eqnarray}
$\sigma_i(i=1,2,3)$ being the usual pseudospin operators and $\lambda$ a coupling constant. We denote by $|g\rangle$ and $|e\rangle$  the two atomic eigenstates with energies $-{\omega_0/2}$ and $+{\omega_0/2}$ respectively, and assume the field to be initially in its vacuum state. Finally, $\omega_0$
takes also into account any direct modification of the atomic transition frequency due to the atomic acceleration.

In order to obtain the resonance interaction between the two accelerated atoms, we follow a procedure due to Dalibard \emph{at al.} \cite{DDC82,DDC84}, allowing to separate vacuum fluctuations and radiation reaction contributions to the energy shift due to the atom-field coupling. This procedure consists of solving the Heisenberg equations of motion for the dynamical variables of the atoms and the field, and separating the solutions into free and source parts. Then, the time evolution of a generic atomic observable can be split in the sum of free and source contributions. Similarly as in \cite{AM94, RLMNSZP16}, we find that the time evolution of a generic atomic observable, pertaining, for example, to atom $A$, can be described in terms of two effective Hamiltonians, $(H_A^{eff})_{vf}$ (related to vacuum field fluctuations) and $(H_A^{eff})_{rr}$ (related to the radiation reaction field). We get
\begin{eqnarray}
&\ & (H_A^{eff})_{vf}=-i{\lambda^2\/4}\int^{\tau}_{\tau_0}d\tau'\langle0\vert\{\phi^f(x_A(\tau)),\phi^f(x_A(\tau'))\}\vert 0\rangle\nonumber\\
&\ &\times [\sigma^f_{2,A}(\tau),\sigma^f_{2,A}(\tau')]\;,
\label{eq:10}
\end{eqnarray}
where $\mid 0 \rangle$ denotes the vacuum state of the field, $\tau_0 \rightarrow -\infty$ is an initial time and $\phi^f(x(\tau))$ is the free part of the field operator which can be expressed as
\begin{eqnarray}
&\ &\phi^f(x(\tau))=\int^{\infty}_0d\omega\int^{\infty}_{-\infty} dk_y\int^{\infty}_{-\infty} dk_z\nonumber\\
&\ &\times
[\;b_{\omega,k_y,k_z}(\tau_0)v_{\omega,k_y,k_z}(\tau,\vec{x})+b^{\dag}_{\omega,k_y,k_z}(\tau_0)v^{\star}_{\omega,k_y,k_z}(\tau,\vec{x})] ,
\label{eq:11}
\end{eqnarray}
with
\beq
v_{\omega,k_y,k_z}(\tau,\vec{x})=e^{-i\omega\tau}v_{\omega,k_y,k_z}(\vec{x})\;,
\label{eq:12}
\eeq
We also have
\beq
\sigma^f_{2}(\tau)={i\/2}(\sigma^f_{-}(\tau)-\sigma^f_{+}(\tau)) ,
\label{eq:13}
\eeq
with
\beq
\sigma^f_{\pm}(\tau)=\sigma_{\pm}(\tau_0)e^{\pm i\omega_0(\tau-\tau_0)}\;,
\label{eq:14}
\eeq
where $\sigma_+=|e\rangle\langle g|$ and $\sigma_-=|g\rangle\langle e|$.

Similarly, $(H_A^{eff})_{rr}$ is
\begin{eqnarray}
&\ & (H_A^{eff})_{rr}=-i{\lambda^2\/4}\int^{\tau}_{\tau_0}d\tau'\langle0|[\phi^f(x_A(\tau)),\phi^f(x_A(\tau'))]|0\rangle\nonumber\\
&\ &\times\{\sigma^f_{2,A}(\tau),\sigma^f_{2,A}(\tau')\}\nonumber\\
&\ &-i{\lambda^2\/4}\int^{\tau}_{\tau_0}d\tau'\langle0|[\phi^f(x_A(\tau)),\phi^f(x_B(\tau'))]|0\rangle\nonumber\\
&\ &\times\{\sigma^f_{2,A}(\tau),\sigma^f_{2,B}(\tau')\}\;.
\label{eq:15}
\eea

Similar expressions are obtained for the effective Hamiltonian governing the time evolution of atom $B$.

As shown by the expressions above, the effective Hamiltonian (\ref{eq:10}) and the first term in Eq.~(\ref{eq:15}), do not depend on the presence of the atom $B$; thus, they contribute only to the single-atom Lamb shift. On the contrary, the second term in Eq.~(\ref{eq:15}) depends on both atoms $A$ and $B$, and thus, it is the only one relevant for the resonance interatomic energy we are calculating.  The resonance interaction energy is then obtained by taking the expectation value of the effective Hamiltonians $(H_{A/B}^{eff})_{rr}$  on one of the two correlated (symmetric or antisymmetric) atomic states,
\beq
|\psi_{\pm}\rangle={1\/\sqrt{2}}(|g_A,e_B\rangle\pm|e_A,g_B\rangle).
\label{eq:16}
\eeq
Taking into account only the terms depending on the interatomic distance, the relevant energy shift is
\begin{eqnarray}
\delta E = \langle \psi_{\pm}\vert (H_A^{eff})_{rr}+(H_B^{eff})_{rr}\vert \psi_{\pm}\rangle.
\label{eq:17}
\end{eqnarray}

We get
\begin{eqnarray}
&\ &\delta E=-i{\lambda^2}\int^{\tau}_{\tau_0}d\tau'\chi^F((x_A(\tau),x_B(\tau'))C_{A,B}(\tau,\tau')\nonumber\\
&\ &+(A\rightleftharpoons B\;\text{term})
\label{eq:18}
\end{eqnarray}
where we have introduced the linear susceptibility of the field, $\chi^F((x_A(\tau),x_B(\tau'))$, defined as
\beq
\chi^F((x_A(\tau),x_B(\tau'))={1\/2}\langle0|[\phi^f(x_A(\tau)),\phi^f(x_B(\tau'))]|0\rangle\;,
\label{eq:19}
\eeq
and the symmetric statistical function of the atoms, $C_{A,B}(\tau,\tau')$,
\beq
C_{A,B}(\tau,\tau')={1\/2}\langle\psi_{\pm}|\{\sigma^f_{2,A}(\tau),\sigma^f_{2,B}(\tau')\}|\psi_{\pm}\rangle\;,
\label{eq:20}
\eeq
which can be further simplified by the use of Eqs.~(\ref{eq:12}), (\ref{eq:14}) and (\ref{eq:16}), obtaining
\beq
C_{A,B}(\tau,\tau')=\pm{1\/8}(e^{i\omega_0(\tau-\tau')}+e^{-i\omega_0(\tau-\tau')})\;.
\label{eq:21}
\eeq
Hereafter, ``$\pm$'' corresponds to the symmetric or antisymmetric state $|\psi_{\pm}\rangle$ respectively.

The procedure outlined above is general, and we can now apply it to calculate the resonance interatomic energy in a coaccelerated frame. Using the Rindler coordinates, the trajectories of the two atoms described by Eq.~(\ref{eq:1}) are now given by
\bea
\tau_A=\tau\;,\quad\;\xi_A=y_A=0\;,\quad\;z_A=z_1\;;\nn\\
\tau_B=\tau\;,\quad\;\xi_B=y_B=0\;,\quad\;z_B=z_2\;.
\label{eq:22}
\eea
Using (\ref{eq:22}) in Eqs.~(\ref{eq:7}), (\ref{eq:11}), (\ref{eq:12}) and (\ref{eq:19}), and after some algebraic calculations, we obtain an explicit expression for the linear susceptibility of the field,
\begin{eqnarray}
&\ &\chi^F(x_A(\tau),x_B(\tau'))=-{1\/8\pi^2}{1\/{z\sqrt{N(z,a)}}}\int^{\infty}_0d\omega g(\omega,z,a)\nonumber\\
&\ &\times(e^{i\omega(\tau-\tau')}-e^{-i\omega(\tau-\tau')})\;,
\label{eq:24}
\end{eqnarray}
where we have defined $z=z_A-z_B$, $N(z,a)=1+{1\/4}a^2z^2$, and $g(\omega,z,a)=\sin({2\omega\/a}\sinh^{-1}({az\/2}))$. A comparison with the result obtained in a local inertial frame [see Eq.~(13) in \cite{RLMNSZP16}], shows the complete equivalence of the results in the locally inertial and Rindler frames. Using Eqs. (\ref{eq:24}) and (\ref{eq:21}) into Eq.(\ref{eq:18}), after some algebra, we obtain the resonant interatomic energy between the two atoms in the following form
\beq
\delta E=\mp{\lambda^2\/16\pi}{1\/z\sqrt{N(z,a)}}\cos\biggl({2\omega_0\/a}\sinh\biggl({az\/2}\biggr)\biggr)\;.
\label{eq:25}
\eeq

This expression gives the resonance interaction between two atoms in the coaccelerated frame; as already mentioned, it coincides with the expression obtained  for the resonance interaction between two atoms uniformly accelerated in the Minkowski space in a locally inertial frame [see Eq.~(14) in Ref.~\cite{RLMNSZP16}].
Thus, although the Minkowski and Rindler vacua are not equivalent due to the Unruh effect, the resonance interaction between two uniformly accelerated atoms {\em seen} by an inertial observer comoving with the two atoms (Minkowski observer) coincides with that {\em seen} by an observer in a coaccelerated frame (Rindler observer), without any additional assumption of an  Unruh temperature for the field in the coaccelerated frame.
This equivalence between the two different (locally inertial and coaccelerated) frames, as far as the resonance interaction is concerned, is related to the intrinsic {\em nonthermal} nature of the resonance interaction, which is exclusively related to the radiation reaction field.
Thus, the assumption of an Unruh temperature for the field in a coaccelerated frame, is not necessary in the present case for a complete equivalence between locally inertial and coaccelerated point of views; this result should be compared with previous results concerning other radiative processes, such as the Lamb shift or the spontaneous emission of a uniformly accelerated atom, where a fully equivalence between the two points of view requires to assume an Unruh temperature for the field in the Rindler frame.

\section{The electromagnetic field case}
\label{sec:3}

We now suppose that the two atoms interact with the quantum electromagnetic field in the vacuum state and are prepared in one of the correlated states (\ref{eq:16}). We use the same procedure of Sec. II for separating the vacuum fluctuations and radiation reaction contributions to the resonance interaction between the two atoms. We first introduce some relevant properties of the quantum electromagnetic field in the Rindler spacetime. Research works dealing with the quantization of electromagnetic field in the Rindler spacetime exploit different gauge conditions \cite{CD77, HMS92a, HMS92b, Moretti97, LOY08}. Here we use the scheme proposed in Ref.~\cite{LOY08}, where the electromagnetic field is quantized under the Weyl gauge. This will make our calculations simpler than with other gauge conditions.  We specialize it to the 4-dimension case.

Under the Weyl gauge, the 0th component of the vector potential is chosen to vanish, i.e.,
\beq
A_0(\tau,\vec{x})=0\;,
\label{eq:26}
\eeq
and the Gauss law reads
\beq
\partial_i(\sqrt{|g|}g^{00}\partial_0A^i)+\sqrt{|g|}j^{0}=0\;,
\label{eq:27}
\eeq
where $i=1,2,3$, $j^{0}$ is the $0$th component of the $4-$vector $j^{\nu}$, and $g^{\mu \nu}$ is the metric tensor. Decomposing the components of the vector potential into the transverse part, $\hat{A}^i$, and the longitudinal part, $A^i_l$,
\beq
A^i=\hat{A}^i+A^i_l\;,
\label{eq:28}
\eeq
with
\beq
\partial_i\hat{A}^i=0\;,\quad\bigtriangledown\times\vec{A}_l=0\;,
\label{eq:29}
\eeq
it is easy  to deduce from Eq.~(\ref{eq:27}) that
\beq
\partial_0A^i_l=0 ,
\label{eq:30}
\eeq
when no external current is present. As a result, the component of the conjugate momentum
\beq
\Pi^i=-\sqrt{|g|}g^{00}\partial_0A^i
\label{eq:31}
\eeq
reduces to
\beq
\Pi^i=-\partial_0\hat{A}^i=E^i ,
\label{eq:32}
\eeq
where $E^i$ denotes the $i$th component of the electric field. So under the Weyl gauge, when no external source appears, the electric field is determined by the transverse part of the vector potential.

The transverse part of the $1$st component of the vector potential can be expanded in terms of a complete set of normal modes as~\cite{LOY08}
\begin{eqnarray}
&\ & \hat{A}_1(\tau,\xi,\vec{x}_{\perp})={1\/2\pi}\int{d\omega d\vec{k}_{\perp}\/{\sqrt{2\omega}}}{a^2\/\omega k_{\perp}}\mathcal{Z}^2k_{i{\omega\/a}}(\mathcal{Z})\nonumber\\
&\ &\times
[a^1_{\omega,\vec{k}_{\perp}}(\tau)e^{i\vec{k}_{\perp}\cdot\vec{x}_{\perp}}+H.c.]\;,
\label{eq:33}
\end{eqnarray}
where $\vec{x}_{\perp}=(y,z)$, $\vec{k}_{\perp}=(k_y,k_z)$, H.c. means ``Hermitian conjugate'', and we have defined
\begin{eqnarray}
\mathcal{Z}={k_{\perp}\/a}e^{a\xi}\;,\quad\;k_{i{\omega\/a}}(\mathcal{Z})={1\/\pi}\sqrt{{2\omega\/a}\sinh\biggl({\pi\omega\/a}\biggr)}K_{i{\omega\/a}}(\mathcal{Z})\; .
\label{eq:34}
\end{eqnarray}
The expansion of the two other components of the vector potential is
\begin{eqnarray}
&\ &\hat{A}_I(\tau,\xi,\vec{x}_{\perp})={1\/2\pi}\int{d\omega d\vec{k}_{\perp}\/{\sqrt{2\omega}}}
\biggl\{\biggl[e_I(\vec{k}_{\perp})a^2_{\omega,\vec{k}_{\perp}}(\tau)\biggr.\biggr.\nonumber\\
&\ & \biggl.\biggl. +i{ak_{I}\/\omega k_{\perp}}a^1_{\omega,\vec{k}_{\perp}}(\tau)\mathcal{Z}{d\/d\mathcal{Z}}\biggr]k_{i{\omega\/a}}(\mathcal{Z})e^{i\vec{k}_{\perp}\cdot\vec{x}_{\perp}}+H.c.\biggr\}\;,
\label{eq:35}
\end{eqnarray}
($I=2,3$). The commutation relations are
\begin{eqnarray}
[a^i_{\omega,\vec{k}_{\perp}}(\tau),a^{\dag j}_{\omega',\vec{k}'_{\perp}}(\tau)]=\delta_{ij}\delta(\omega-\omega')\delta(\vec{k}_{\perp}-\vec{k}'_{\perp}),
\label{eq:36}
\end{eqnarray}
$(i,j=1,2)$, and the polarization vector $\hat{e}(\vec{k}_{\perp})$ satisfies the following relations
\begin{eqnarray}
&\ &\sum^{3}_{I=2}k_I\hat{e}_I(\vec{k}_{\perp})=0,\,\,
\sum^{3}_{I=2}\hat{e}^2_I(\vec{k}_{\perp})=1,\nonumber\\\
&\ &\hat{e}_I(\vec{k}_{\perp})\hat{e}_J(\vec{k}_{\perp})=\delta_{IJ}-{\hat{k}_I \hat{k}_J\/k^2_{\perp}} ,
\label{eq:37}
\end{eqnarray}
with $J=2,3$. According to the relations above, we can choose the polarization vectors in the following form
\bea
\hat{e}(\vec{k}_{\perp})=\biggl(-{k_z\/k_{\perp}},{k_y\/k_{\perp}}\biggr)\;,\quad \text{or}\quad\;\hat{e}(\vec{k}_{\perp})=\biggl({k_z\/k_{\perp}},-{k_y\/k_{\perp}}\biggr)\;.
\label{eq:38}
\eea
The Hamiltonian of the electromagnetic field, in terms of the creation and annihilation operators, is
\beq
H_F(\tau)=\int d\omega\int d\vec{k}_{\perp}\;\omega\sum^{2}_{i=1}a^{\dag i}_{\omega,\vec{k}_{\perp}}(\tau)\;a^i_{\omega,\vec{k}_{\perp}}(\tau)\;.
\label{eq:39}
\eeq

We now consider the two uniformly accelerated identical atoms prepared in one of the correlated states (\ref{eq:16}) in a coaccelerated frame and interacting with the quantum electromagnetic field. The total Hamiltonian of the interacting system in the multipolar coupling scheme and in the dipole approximation has thus the form
\begin{eqnarray}
H(\tau)&=&H_A(\tau)+H_B(\tau)+H_F(\tau) \nonumber\\
&-&\vec{\mu}_A(\tau)\cdot\vec{E}(x_A(\tau)) -\vec{\mu}_B(\tau)\cdot\vec{E}(x_B(\tau)),
\label{eq:40}
\end{eqnarray}
where $\vec{\mu}=e\vec{r}$ is the atomic dipole moment operator and $\vec{E}(x(\tau))$ the electric field operator.

As before, the resonance interaction energy between the two atoms can be obtained as
\begin{eqnarray}
\delta E&=&-ie^2\int^{\tau}_{\tau_0}d\tau'\chi^F_{ij}(x_A(\tau),x_B(\tau'))\;C^{A,B}_{ij}(\tau,\tau\textquoteright)\nonumber\\
&+&(A\rightleftharpoons B)
\label{eq:41}
\end{eqnarray}
where $\chi^F_{ij}(x_A(\tau),x_B(\tau'))$ is the linear susceptibility of the quantum electromagnetic field
\beq
\chi^F_{ij}(x_A(\tau),x_B(\tau'))={1\/2}\langle0|[E_i(x_A(\tau)),E_j(x_B(\tau\textquoteright))]|0\rangle\;,
\label{eq:42}
\eeq
and $C^{A,B}_{ij}(\tau,\tau')$ is the atomic symmetric statistical function
\beq
C^{A,B}_{ij}(\tau,\tau')={1\/2}\langle\psi_{\pm}|\{r^{Af}_i(\tau),r^{Bf}_j(\tau')\}|\psi_{\pm}\rangle\;,
\label{eq:43}
\eeq
with
\beq
r^{Af}_i(\tau)=\sum_{mn}(r_i)_{mn}\sigma^A_{mn}(\tau_0)e^{-i\omega_{mn}(\tau-\tau_0)}\; .
\label{eq:44}
\eeq
A similar expression is obtained for $r^{Bf}_i(\tau)$. From Eq.~(\ref{eq:44}) and considering two-level atoms,  we find
\beq
C^{A,B}_{ij}(\tau,\tau')=\pm{1\/2}(r^A_{ge})_i(r^B_{eg})_j(e^{i\omega_0(\tau-\tau')}+e^{-i\omega_0(\tau-\tau')})\;.
\label{eq:45}
\eeq

Now, substituting Eqs.~(\ref{eq:32})-(\ref{eq:38}) into Eq.~(\ref{eq:42}),
after some calculation we obtain the following expression of the field linear susceptibility
\begin{widetext}
\begin{eqnarray}
&\ &\chi^F_{ij}(x_A(\tau),x_B(\tau'))={1\/8\pi^2}\int^{\infty}_0d\omega(e^{-i\omega(\tau-\tau')}-e^{i\omega(\tau-\tau')})\nn\\&\ &
\times\biggl[f_{ij}(a,z,\omega)\cos\biggl({2\omega\/a}\sinh\biggl({az\/2}\biggr)\biggr)
+g_{ij}(a,z,\omega)\sin\biggl({2\omega\/a}\sinh\biggl({az\/2}\biggr)\biggr)\biggr]\;,\nn\\
\label{eq:46}
\end{eqnarray}
\end{widetext}
where we have introduced the functions $f_{ij}(a,z,\omega)$ and $g_{ij}(a,z,\omega)$, whose components are given by
\begin{eqnarray}
&\ &    f_{xx}(a,z,\omega)={\omega(1+a^2z^2)\/z^2N^2(z,a)}\;,\\
&\ &    f_{yy}(a,z,\omega)={\omega(1+{1\/2}a^2z^2)\/z^2N(z,a)}\;,\\
&\ &    f_{zz}(a,z,\omega)=-{2\omega(1+{1\/8}a^2z^2+{1\/16}a^4z^4)\/z^2N^2(z,a)}\;,\\
&\ &    f_{xz}(a,z,\omega)=-f_{zx}(a,z,\omega)={a\omega(1-{1\/2}a^2z^2)\/2zN^2(z,a)}\;,
\label{eq:47}
\end{eqnarray}

\begin{widetext}
\bea
&\ &    g_{xx}(a,z,\omega)=-{1+{1\/4}a^2z^2(2+a^2z^2)-\omega^2z^2(1+{1\/4}a^2z^2)\/z^3N^{5\/2}(z,a)}\;,\\
&\ &    g_{yy}(a,z,\omega)=-{1-\omega^2z^2(1+{1\/4}a^2z^2)\/z^3N^{3\/2}(z,a)}\;,\\
 &\ &   g_{zz}(a,z,\omega)={2[(1+{5\/8}a^2z^2)-{1\/8}a^2\omega^2z^4(1+{1\/4}a^2z^2)]\/z^3N^{5\/2}(z,a)}\;,\\
&\ &    g_{xz}(a,z,\omega)=-g_{zx}(a,z,\omega)=-{a[(1+a^2z^2)+\omega^2z^2(1+{1\/4}a^2z^2)]\/2z^2N^{5\/2}(z,a)}\;.
\label{eq:48}
\eea

Finally, substitution of Eqs.~(\ref{eq:46}) and (\ref{eq:45}) into Eq.~(\ref{eq:41}), after some algebra, yields

\bea
\delta E&=&\pm{{\delta_{ij}(\mu^A_{ge})_i(\mu^B_{eg})_j}\/{4\pi}}\times
\biggl[f_{ij}(a,z,\omega_0)\sin\biggl({2\omega_0\/a}\sinh^{-1}\biggl({az\/2}\biggr)\biggr)
-g_{ij}(a,z,\omega_0)\cos\biggl({2\omega_0\/a}\sinh^{-1}\biggl({az\/2}\biggr)\biggr)\biggr]\nn\\
&\pm&{1\/{4\pi}}[{(\mu^A_{ge})_x(\mu^B_{eg})_z}-{(\mu^A_{ge})_z(\mu^B_{eg})_x}]\times
\biggl[f_{xz}(a,z,\omega_0)\sin\biggl({2\omega_0\/a}\sinh^{-1}\biggl({az\/2}\biggr)\biggr) \nonumber \\
&-&g_{xz}(a,z,\omega_0)\cos\biggl({2\omega_0\/a}\sinh^{-1}\biggl({az\/2}\biggr)\biggr)\biggr]\
\label{eq:49}
\eea
\end{widetext}
(the Einstein sum rule for the repeated indices $ij$ has been adopted).

The expression (\ref{eq:49}) gives the resonance interaction between two atoms in the coaccelerated frame and it is valid for any distance between the atoms (beyond distances of overlap of their electronic wave function). It coincides with that obtained in a locally inertial frame [see Eq. (27) in Ref.~\cite{RLMNSZP16}], analogously to the scalar field case discussed in the previous section. Although the same results are obtained in the instantaneously inertial frame and in the coaccelerated frame for both the scalar and the electromagnetic field cases, one should keep in mind that these two frames are different. In order to obtain the interaction energy in the comoving frame, infinite instantaneously inertial observers are necessary. At any instant, the observer measures an interaction energy which sum up to the interaction energy given by Eqs.~(\ref{eq:25}) and (\ref{eq:49}). However, one coaccelerated observer is enough to get the interaction energy, if the observation time interval is long enough.

As discussed before, the equivalence of the results in the coaccelerated and comoving frames stems from the intrinsically {\em nonthermal} property of the linear field susceptibility in both locally inertial and coaccelerated frames. Thus, similarly to the scalar field case, also in the case of atoms interacting with the quantum electromagnetic field in the vacuum state, we find a complete agreement between the results here obtained in the coaccelerated frame and the known results obtained in a locally inertial frame, without any extra condition on the Unruh temperature. In this sense, from the point of view of the resonance interatomic interaction, a coaccelerated frame is equivalent to a locally inertial frame.

It is worth pointing out that, although the resonance interatomic energy is calculated assuming the field (in the coaccelerated frame) in its vacuum state, the results obtained (for the comoving and Rindler frames) are valid even if we assume the field in a thermal state at arbitrary temperature, as the linear susceptibility function is independent of the temperature. This conclusion is basically related to the intrinsic
nonthermal nature of the resonance interaction. The situation is totally different when we consider only one atom coupled to a massless scalar field~\cite{ZY07} or to the electromagnetic field~\cite{ZY08}; in this case, it has been shown that the average rate of change of the atomic energy in a coaccelerated frame is equivalent to the counterpart in a locally inertial frame, if the field in the coaccelerated frame is assumed to be in a thermal state at the Unruh temperature, proportional to the acceleration of the system; in this case, the average rate of change of the atomic energy in the coaccelerated frame is temperature-dependent. This behavior can be traced back to the fact that the rate of atomic transition is related to both field fluctuations and radiation reaction contributions and that the contribution of the field fluctuations is always temperature-dependent. Similar results are expected for the Casimir-Polder dispersion interatomic interaction between two uniformly accelerated atoms, where one expects that dispersion interactions in a coaccelerated frame will show an obvious temperature-dependent behavior, related to the field fluctuations contribution. Thus our results clearly show that the resonance interaction has, for the aspects discussed above, a completely distinctive behavior.

\section{Conclusions}
\label{sec:4}

We have investigated the resonance interatomic energy between two uniformly accelerated atoms, one in an excited state with the other in the ground state and prepared in a correlated state, from the viewpoint of a coaccelerated observer. We have considered both the cases of atoms interacting with a massless scalar field and the electromagnetic field, in the vacuum state. We have separated the contributions of vacuum field fluctuations and radiation reaction to the resonance interaction between the two atoms and discussed that only the radiation reaction field contributes to the resonance interaction, which is thus independent on the thermal radiation in the space. We have shown that the linear field susceptibility in the coaccelerated frame coincides  with that obtained in a local inertial frame. Thus, in both cases considered, we obtain in the coaccelerated frame  the same expressions obtained in a locally inertial frame, without any additional assumption of the Unruh temperature for the quantum fields in the coaccelerated frame (contrarily to other radiative effects for uniformly accelerated atoms such as Lamb shift and spontaneous emission, for example), thus making clearer the relation between acceleration and temperature for accelerated systems.

\begin{acknowledgments}
This work was supported in part by the National Natural Science Foundation of China (NSFC) under Grants No. 11375092 and No. 11405091; the Zhejiang Provincial Natural Science Foundation of China under Grant No. LQ14A050001, China Scholarship Council (CSC), the Research Program of Ningbo University under No. E00829134702, and K.C. Wong Magna Fund in Ningbo University.
Financial support by the Julian Schwinger Foundation, MIUR, University of Palermo, is also gratefully acknowledged.
\end{acknowledgments}

\end{document}